\documentclass[runningheads]{llncs}
\usepackage{graphicx}
\usepackage{lipsum}  
\usepackage{amsmath}
\usepackage{color}
\usepackage{float}
\usepackage{array}
\usepackage{subfigure}
\usepackage{tabularx}
\usepackage{pgfplots}
\usepackage{subfig}
\newcommand{\myparagraph}[1]{\smallskip\noindent\textbf{#1}}

\begin{document}
\title{Attention based CNN-LSTM Network for Pulmonary Embolism Prediction on Chest Computed Tomography Pulmonary Angiograms}
%
\titlerunning{Attention based CNN-LSTM Network for PE Prediction}
%
\author{Sudhir Suman\inst{1} \and
Gagandeep Singh\inst{2} \and
Nicole Sakla\inst{2} \and
Rishabh Gattu\inst{2} \and
Jeremy Green\inst{2}  \and
Tej Phatak\inst{2}   \and
Dimitris Samaras\inst{3}  \and
Prateek Prasanna\inst{4}}

\authorrunning{Suman et al.}
%

\institute{Dept. of Electrical Engineering, Indian Institute of Technology, Bombay, India \and
Dept. of Radiology, Newark Beth Israel Medical Center, NJ, USA  \and
Dept. of Computer Science, Stony Brook University, NY, USA  \and
Dept. of Biomedical Informatics, Stony Brook University, NY, USA  
\email{prateek.prasanna@stonybrook.edu} 
}

\maketitle              
\begin{abstract}
With more than 60,000 deaths annually in the United States, Pulmonary Embolism (PE) is among the most fatal cardiovascular diseases. It is caused by an artery blockage in the lung; confirming its presence is time-consuming and is prone to over-diagnosis. The utilization of automated PE detection systems is critical for diagnostic accuracy and efficiency. In this study we propose a two-stage attention-based CNN-LSTM network for predicting PE, its associated type (chronic, acute) and corresponding location (leftsided, rightsided or central) on computed tomography (CT) examinations. We trained our model on the largest available public Computed Tomography Pulmonary Angiogram PE dataset (RSNA-STR Pulmonary Embolism CT (RSPECT) Dataset, N=7279 CT studies) and tested it on an in-house curated dataset of N=106 studies. Our framework mirrors the radiologic diagnostic process via a multi-slice approach so that the accuracy and pathologic sequela of true pulmonary emboli may be meticulously assessed, enabling physicians to better appraise the morbidity of a PE when present.  Our proposed method outperformed a baseline CNN classifier and a single-stage CNN-LSTM network, achieving an AUC of 0.95 on the test set for detecting the presence of PE in the study.

\end{abstract}
\footnotetext[1]{This work will be presented at MICCAI 2021}
\section{Introduction}

\myparagraph{Clinical Motivation.} 
Pulmonary embolism (PE) is the most common preventable cause of hospital death in the United States with diagnostic delay and diagnostic challenges among the most common etiologies for resultant mortality \cite{friedman2018patient}. The time dependent nature of PE diagnosis is especially critical given that the treatment involves anti-coagulant therapy which must be administered in an attempt to halt thrombus growth. Delayed diagnosis and treatment enables undiagnosed clots to increase in size. The patient’s mortality therefore becomes compounded by not only the resultant ischemic insult to the pulmonary parenchyma but also by the potential development of decompensated right heart failure.  Currently, attempts to reduce diagnostic delays have predominantly focused on the speed of PE symptom identification in the emergency setting and little to no changes made to the imaging diagnosis paradigm through which treatment is ultimately decided \cite{friedman2018patient}.  Through the application of machine learning techniques, the diagnostic delays and errors in PE identification may be mitigated, thereby decreasing patient mortality.

\myparagraph{Technical Motivation.} 
Similar to the neoplastic applications of deep learning paradigms~\cite{sirinukunwattana2016locality,coudray2018classification,bejnordi2017diagnostic}, the utilization of automated systems for PE detection is critical when diagnostic accuracy and efficiency are considered. 
Early works in automated PE diagnosis have mostly relied on traditional feature engineering and pulmonary vessel segmentation to reduce the search space which is quite computationally intensive~\cite{liang2007computer,masutani2002computerized,zhou2005preliminary,ozkan2014novel,park2010multistage}. Furthermore, most of these studies have reported results on small datasets. Recent Convolutional Neural Network-Long Short-Term memory (CNN-LSTM) methods~\cite{rajan2020pi,shi2020automatic} and an end-to-end deep learning model, PENet~\cite{huang2020penet}, have demonstrated promise when diagnosing PE on Computed Tomography Pulmonary Angiogram (CTPA) examinations, but are overall limited in their binary classification of whether a PE is present or not. Methods based on blood clot segmentation \cite{rajan2020pi,yang2019two} for detecting PE are quite computationally-intensive, due to their use of only segmented blood clot regions; other regions which may show pulmonary abnormalities suggestive of clots get ignored.

In this work, we present a two-stage attention-based network consisting of a CNN and a Sequence model (LSTM + Dense) for prediction of PE and its associated characteristics. Current detection algorithms are largely modeled on the premise that every single CT slice is evaluated as an independent unit versus analyzing the relationship between successive CT slices. Our model extrapolates information from this 3D relationship in an attempt to better mimic the human cognitive process when examining a cross-sectional image. 
Due to the fact that PE is often observed in a small subset of 2D slices of a CT volume, the proposed attention module assists the network in locating insightful slices and assigns them higher weights in a bag-level feature aggregation mechanism. The slices with a higher attention are more likely to contain important information for PE identification thereby aiding in subsequent PE prediction. Besides, unlike other methods, our framework is trained to identify forms of PE on each CTPA examination slice and other PE attributes such as laterality, chronicity, and the RV/LV (Right Ventricular to Left Ventricular) ratio which can prove clinically useful in determining patients at risk~\cite{lankeit2017always,ghaye2006can}. 
Our contributions can be summarized as follows:
\begin{enumerate}
    \item Our attention based network  provides image-level ($\psi_{image}$) PE prediction for each of the CTPA examination slices, study-level ($\psi_{study}$) prediction of PE in the CTPA volume and other associated PE characteristics.
    \item Multiple Instance Learning (MIL) pooling \cite{ilse2018attention} is used as an attention mechanism that provides insight into the contribution of each CTPA examination slice and the aggregate feature representation for study-level prediction.  
    \item Our network is trained on the largest publicly available PE dataset (RSPECT N$>$7000 studies) \cite{colak2021rsna} and its performance has been evaluated on an external test dataset (N=106 studies).
    \item Our network is designed and trained to explicitly obey clinically-defined label hierarchy and avoid making conflicting label predictions.
\end{enumerate}

\section{Proposed Methodology}
Our network architecture comprises of two stages: 1) a CNN classifier to capture the image properties  and study labels and 2) a sequence model for learning inter-slice dependencies. 
The CNN is used to extract features from every slice from the study and the sequence model combines these \textit{spatially dependent} features and captures long-range dependencies to give the network information about the global change around each CT slice.


\begin{figure}[h]
\includegraphics[width=\linewidth]{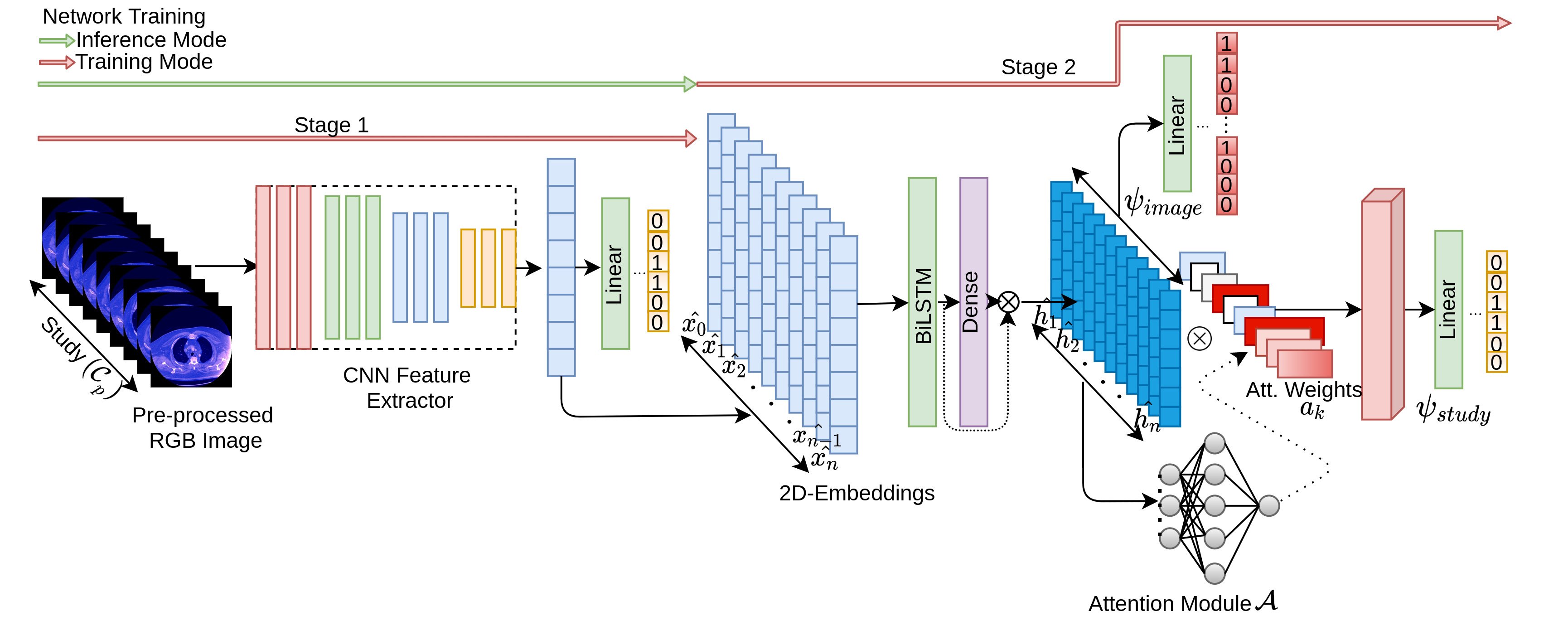}
\caption{\textbf{Overview of our two-stage attention based PE detection model.} The first stage involves training with all slices in study $\mathcal{C}_p$ for all image and study labels. The 2D-embedding $\{{\hat{x}_{1}},{\hat{x}_{2}},..,{\hat{x}_{3}}\}$ from stage 1 after the pooling layer is used for the stage 2 LSTM-based training with an attention module for generating final predictions} 
\label{fig1}
\end{figure}

\subsection{Pre-processing}
CT scans capture tissue radiodensity information. 
Voxel intensities range from -1000 to 3000 Hounsfield Units. Different window levels are used by radiologists to evaluate lung pathologies.
We first convert the single channel images to a 3-channel RGB format~\cite{anam2019novel}, where each channel is a distinct window~\cite{wittram2004ct} corresponding to the lung (window level = -600, width = 1500), PE (window level = 100, width = 700), and mediastinal ranges (window level = 40, width = 400) .
%
The PE and mediastinal windows enable detection of blood clots. The lung window, on the other hand, may show pulmonary abnormalities suggestive of a clot. The constructed 3-channel RGB images were used to train our network.
\subsection{\textit{Stage 1:} CNN Classifier to capture different label properties}
We used a Convolutional Neural Network (CNN) Efficient-Net \cite{tan2019efficientnet} architecture as a backbone to extract features, followed by average pooling and a linear layer to classify each image and study. The pooling layer  reduces the size of features maps after the convolutional layer so that we can make use of all the slices present in the study at once while training the Sequence model in Stage 2. This CNN classifier is trained for multi-label prediction on the  RSPECT~\cite{colak2021rsna} dataset to capture the properties for different study labels. We make use of spatial feature maps generated in Stage 1 to train our Sequence model in Stage 2.
\subsection{\textit{Stage 2:} Sequence Model for learning inter-slice dependencies}
The Sequence model, consisting of a bidirectional long short-term memory (BiLSTM) ~\cite{huang2015bidirectional} and a dense layer, is used to capture long-range dependencies in CT scans. It makes use of the extracted features from study slices using the trained CNN classifier from Stage 1 and passes them through a Bi-LSTM and a dense layer to provide the network additional contextual information about the global changes around the slices. The features extracted using CNN-LSTM network capture spatio-temporal information in the CT scan volume. The `temporal' aspect refers to the global relationship between successive slices. These features are subsequently used for the final prediction. The Sequence model has two heads - one for image ($\psi_{image}$) and another for study ($\psi_{study}$) level predictions. The $\psi_{image}$ head is used for detecting the presence of PE on each slice in the study and the $\psi_{study}$ head provides predictions regarding different characteristics of PE in the given study consisting of hundreds of slices.

\myparagraph{Attention Mechanism.} Features from the Sequence Model's Bi-LSTM and dense layers,  $(\hat{h}_{1},\hat{h}_{2},..,\hat{h}_{n})$, corresponding to slices $\mathcal{C}_{pq}$, $q\in [1, 2, ....n]$  in study volume $\mathcal{C}_p$ are fed into the attention module $\mathcal{A}$ to obtain bag level features for $\mathcal{C}_p$. The aggregated bag-level features are used for global prediction by passing it through the the $\psi_{study}$ classifier. Attention pooling is the weighted average of the slice-wise features present in the study and these attention weights $a_{k}$  are learned from the neural network as
$
a_{k} = \frac{\exp\{w^{T}tanh(V\hat{h}_{k}^{T})\}}{\sum_{j=1}^{n} \exp\{w^{T}tanh(V\hat{h}_{j}^{T})\}}
$, where $w$ and $V$ are learnable parameters and $a_{k}$ denotes the learned attention weight distribution of $\mathcal{A}$. Features from each slice have a corresponding weight, and the bag-level feature set is obtained as $z  = \sum_{k=1}^{n} a_{k}\hat{{h}}_{k}$. The attention module enables the network to locate informative slices present in the study. During feature aggregation, it assigns higher weight to the  more informative slices for final study-level prediction.

\myparagraph{Loss Function.} \label{loss}
We use a custom loss function for training, similar to the performance metrics for the RSNA STR Pulmonary Embolism Detection challenge \cite{challenge}.
The study-level weighted log loss is defined as
$$
L_{ij} = -w_{j}*[y_{ij}*log(p_{ij}) + (1-y_{ij})*log(1-p_{ij})],
$$
 where $i$ and $j$ denote the study and label respectively, $y_{ij}$ is ground truth label \{0,1\} and $p_{ij}$ is prediction probability of label $j$ for study $i$ that $y_{ij}$ = 1. $w_{j}$ signifies the weight of label $j$.
Similarly, the image-level weighted log loss is defined as
$$
L_{ik} = -[w*\frac{\sum_{k=1}^{n_{i}}y_{ik}}{n_{i}}]*[y_{ik}*log(p_{ik}) + (1-y_{ik})*log(1-p_{ik})]
$$
Here $i$ is the study and $k$ = $1,2,3..,n_i$, where $n_i$ is the total number of images in study i, $y_{ik}$ is ground truth \{0,1\} and  $p_{ik}$ is prediction probability (for presence of PE) on image $k$ in study $i$. The total loss of study $i$ is the sum of the image-level and study-level loss divided by the total weights, given by
$$
L_{i} = \frac{\sum_{k=1}^{n_{i}}L_{ik} + \sum_{j=1}^{n}L_{ij}}{\sum_{j=1}^{n}w_{j} + w*\sum_{k=1}^{n_{i}}y_{ik}}.
$$
\begin{figure}
\centering
\includegraphics[width=\linewidth]{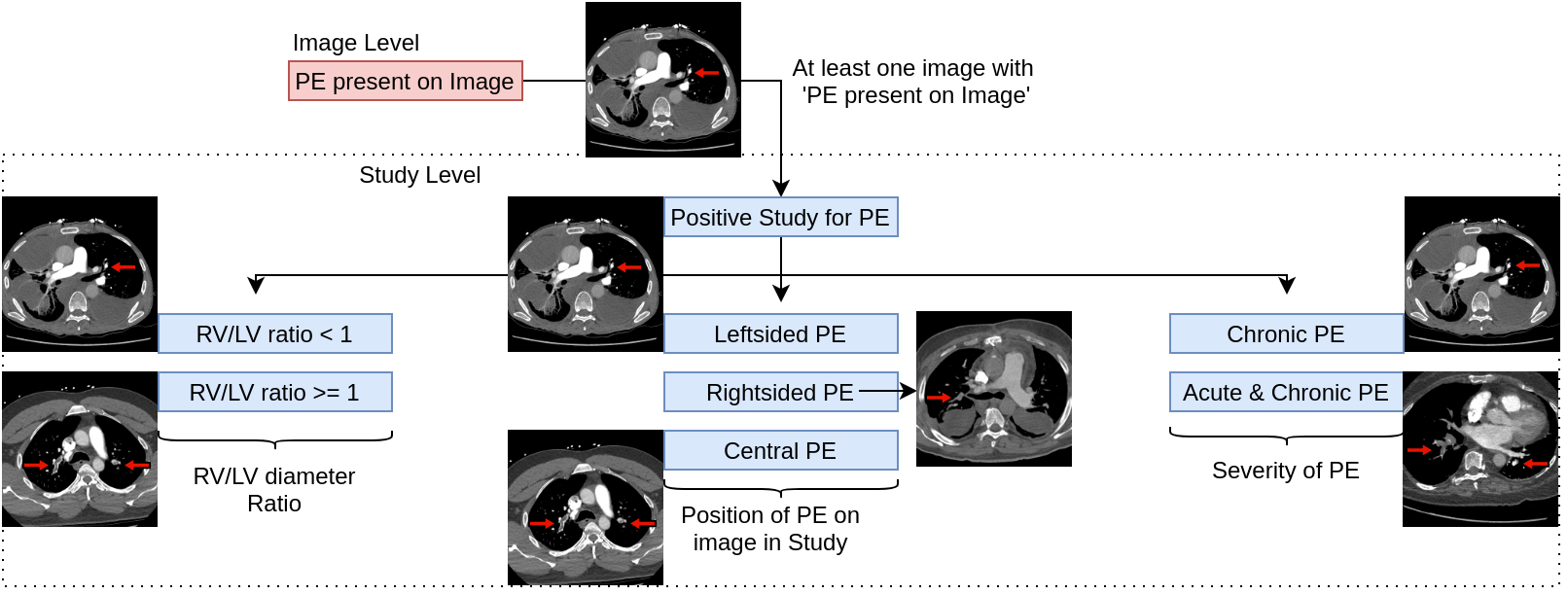}
\caption{Flowchart outlining relationships between labels} \label{fig1}
\end{figure}
\myparagraph{Label Consistency.}
\label{consistenecy}
Our model ensures logical consistency of the predicted labels using constraint-based modified activation function. According to the rules in~\cite{challenge}, a label is considered as predicted if it is assigned a probability ($p$) of more than 0.5. Label consistency rules of PE are as follows:
\begin{itemize}
    \item If any image in a study is predicted positive for the presence of PE
    \begin{itemize} 
        \item At least one of the three labels, i.e, \textit{Right Sided, Left Sided or Central} PE must be assigned a probability greater than 0.5.
        \item Only one of \textit{RV/LV $<$ 1 or  RV/LV $\geq$ 1} will have $p > 0.5$ and one of it must be present if at least one image is found positive for PE in study.
        \item Both \textit{Chronic \& Acute} and \textit{Chronic} cannot be predicted positive at the same time , i.e, only one can have $p > 0.5$.
    \end{itemize}
    \item If no image in the study volume is found positive for presence of PE, then
    \begin{itemize}
        \item Either \textit{indeterminate} or \textit{negative for PE} labels must have a $p > 0.5$; both cannot have a $p > 0.5$.
        \item All positive study related labels, i.e, \textit{right sided PE, left sided PE, RV/LV $<$ 1, RV/LV  $\geq$ 1, central PE}  must have a $p < 0.5$.
    \end{itemize}
\end{itemize}

\begin{table}
\centering
\caption{Data distribution in training, validation and test sets.}\label{tab:my_label}
\begin{tabular}{l|l|l|l}
\hline
labels & Train  & Val  & Test \\
\hline
     \# studies & 5824 & 1455 & 106 \\
     \# studies with positive PE   & 1878 & 490 & 56  \\
    \# study with $RV/LV \geq 1$  & 754 & 186 & 43  \\
    \# study with $RV/LV < 1$  & 1878 & 490 & 63  \\
    \# study with leftsided PE  & 1242 & 302 & 51  \\
    \# study with rightsided PE  & 1486 & 389 & 53  \\
    \# study with central PE  & 318 & 83 & 43  \\
    \# study with Chronic PE  & 220 & 72 & 2  \\
    \# study with Acute\&Chronic PE  & 116 & 29 & 0  \\
    \hline
    \# slice  & 1431318 & 359276 & 87764  \\
    \# slice with PE & 77453 & 19087 & $-$ \\
\hline
\end{tabular}
\end{table}

\section{Experimental Design}
\subsection{Dataset Description}
\myparagraph{Training Dataset.}
We use the largest publicly available annotated PE dataset  (the RSNA Pulmonary Embolism CT, RSPECT dataset), comprising more than 12,000 CT studies~\cite{colak2021rsna}. This dataset, composed of CTPA scans and associated ground truth annotations has been contributed to by five international research centers and labeled by a group of more than 80 expert thoracic radiologists. There are multiple labels for each study and one label for each slice present in the study. The flowchart outlining the relationships between different labels is shown in Figure~\ref{fig1}. 1,790,594 slices corresponding to 7279 studies have a complete set of ground truth labels available. 5824 of these studies were randomly selected as the training ($D_1$) and 1455 studies as the validation  ($D_2$) cohorts. 

\myparagraph{Independent test set.} We retrospectively collected data from 106 patients ($D_3$) who received CTPA studies under the PE protocol (GE Revolution CT, GE Lightspeed VCT, GE Lightspeed 16, Phillips Brilliance 64) at $Blinded Institute X$. 
The images were anonymized in accordance with Health Insurance Portability and Accountability Act and our Institutional Review Board guidelines. 
Both male and female adult patients (over 18 years) were selected with random sampling utilized in order to obtain similar numbers of positive versus negative PE cases (56 positive and 50 negative). CTPA studies with extensive artifacts (i.e. beam hardening, motion artifact, over/under exposure) were excluded. The ground truth study-level labels were provided by three expert readers  working in consensus. Two readers ($>$15 yr and $>$10 yr exp) are board certified radiologists and the third third reader ($>$3 yr exp) is a radiology resident.
\subsection{Framework Implementation} 
We trained our PE network in two stages. First, we trained the CNN classifier using all pre-processed 3-channel slices in $D_1$, with both image and study-level labels as targets. Each batch consists of slices from the same study and the final study-level prediction was based on the prediction with maximum confidence. This multitask learning stage not only captures the presence of PE in slices but also learns other PE properties based on different study-level targets. We extracted the features of each slice from study $\mathcal{C}_p$ using the trained CNN network and map it to a 2-D embedding $\{{\hat{x}_{1}},{\hat{x}_{2}},..,{\hat{x}_{3}}\}$ using global average pooling .
To exploit the sequential inter-slice spatial relationship and learn global intensity variations, we passed the embedding in order of their captured time points to the LSTM layer. The extracted features from the LSTM module passes through the attention module to yield bag-level features for the study volume $\mathcal{C}_p$, which is subsequently fed to the final study-level classifier($\psi_{study}$); features from the LSTM are directly provided to the image-level classifier($\psi_{image}$) for detecting the presence of PE on individual slices. Both stages were trained with the Adam optimizer ~\cite{kingma2014adam} and the custom loss function discussed in section~\ref{loss}. The model with the lowest loss value on $D_2$ was chosen for evaluation on $D_3$. Both stages were implemented in PyTorch 1.4 and trained using one NVIDIA Tesla T4 Google Cloud Instance GPU.
\section{Results}
\begin{figure}[h]
\centering
\includegraphics[width=\linewidth]{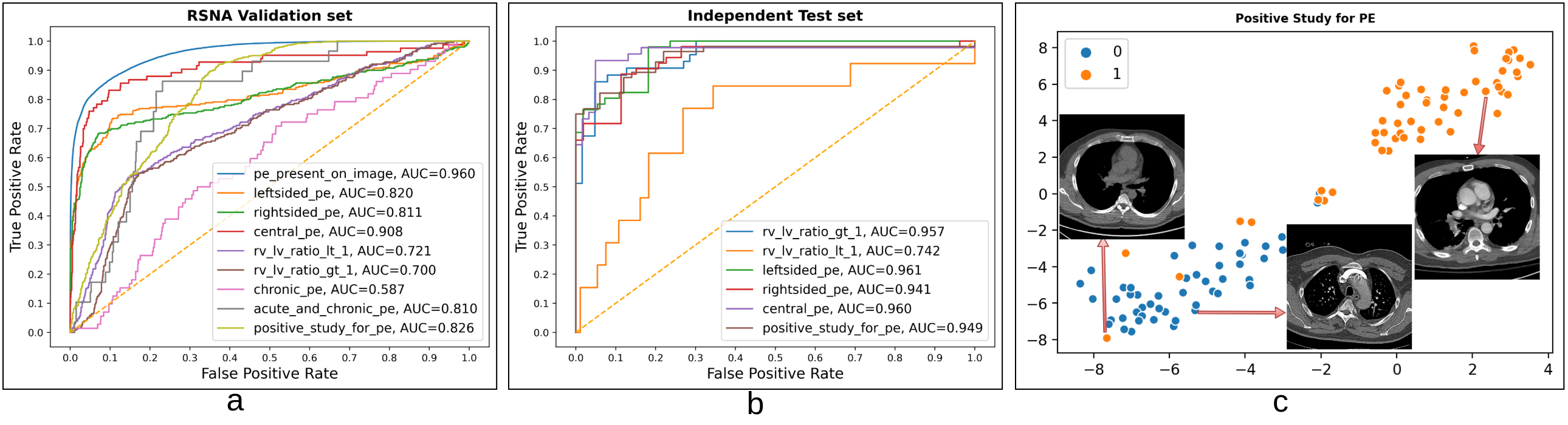}
\caption{(a,b) ROC curves for different image and study-level labels on the validation and independent test set. (c) 2D scatter plot representation using t-SNE of bag level features from Attention Module corresponding to studies in $D3$}. 
\label{tab:auc}
\end{figure}
We used area under the receiver operating characteristic curve (AUC) as our performance metric. Our proposed method was compared with a CNN classifier (Baseline 1) and a CNN-LSTM network without the attention module (Baseline 2). The AUCs of Baseline 1, Baseline 2, and our method were 0.5, 0.94, and 0.95, respectively. The corresponding accuracies were 0.74, 0.65, and 0.88, respectively. 
The performance curves of our best model for different images and study labels on $D_2$ and $D_3$ are shown in Figure \ref{tab:auc}. Our proposed network achieved an AUC of 0.82 for detecting PE on $D_2$ and 0.95 on the $D_3$. CT volumes in $D_3$ had different slice thickness and higher number of slices per study as compared to those in $D_1$ and $D_2$, which potentially provides more contextual information to the model. 

\begin{figure}[h]
\centering
\includegraphics[width=\linewidth]{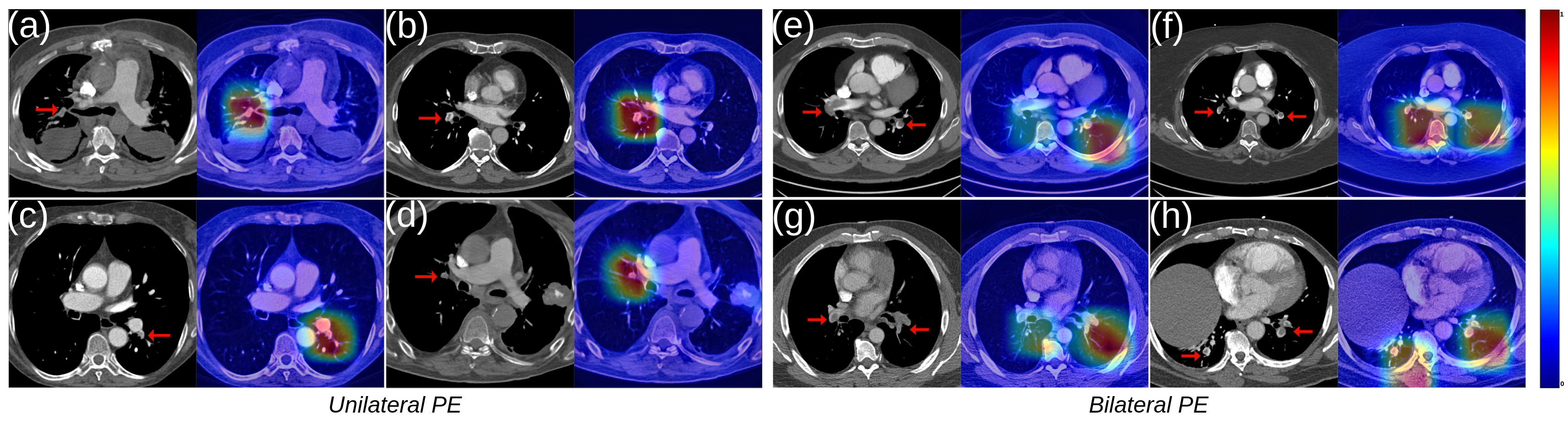}
\caption{\textbf{Visualization of class activation maps.}  CTPA slices (a-d) demonstrate unilateral PE marked by red arrows. High intensity regions in the CAMs represent the precise location of PE. (a, b, d) show right-sided lobar/segmental PE. (c) shows lower segmental PE. Slices (e-h) demonstrate bilateral PE. (f, h) show lobar/segmental PE and CAMs depict the exact location of PE. In (e, g), the CAMs do not accurately define the maximal intensity on the right-sided central/lobar PE; however, they depict left-sided PE accurately.} 
\label{tab:cam}
\end{figure}

To provide an intuitive understanding into the features and the context learnt by the network, class activation maps (CAMs)~\cite{zhou2016learning} were computed by averaging the feature maps from the final convolutional layer of CNN network. 
As may be observed on a few representative slices in Figure~\ref{tab:cam}, there is a strong activation over the regions surrounding the clots, as verified by our collaborating radiologists (see interpretation in Figure~\ref{tab:cam} caption). This corroborates the clinical characteristics of PE diagnosis. A t-Distributed Stochastic Neighbor Embedding (t-SNE) ~\cite{van2008visualizing} was applied to the attention module's bag level features. The 2D scatter plot visualization is shown in Figure~\ref{tab:auc}.

As may be observed in Figure \ref{tab:auc}, our two-stage CNN-LSTM network with an attention mechanism is able to learn and predict not only presence of PE, but also its associated attributes with a high AUROC. Note that the number of ROC curves are fewer in $D_3$ as compared to $D_2$ in Figure~\ref{tab:auc} since image-level annotations and enough number of studies for \textit{chronic} and \textit{acute \& chronic PE} were not available for $D_3$.
\section{Conclusion}
In this work, we developed an attention based two-stage network for CTPA classification on the largest publicly available PE dataset~\cite{colak2021rsna} and tested its performance on a held out validation and an independent test set from a different clinical site.
We believe the contribution is significant not only due to the large-scale multi-institutional validation, but also because of how the framework is designed to explicitly obey a clinically-defined label hierarchy. Our framework is among the first to identify forms of PE on each CTPA slice as well as other attributes such as laterality, chronicity, and the RV/LV ratio which can prove clinically useful in determining patient risk. Our model is also designed to mimic the human cognitive process when examining cross-sectional scans. Furthermore, the generated CAMs have been validated by radiologists.
Our motivation behind a two-stage 2D CNN-LSTM training approach stems from the GPU memory limitations in the PE challenge~\cite{challenge}. The results of the challenge are also suggestive of the fact that a 2D-CNN with a Sequence model works better than a 3D-CNN for this task. We achieved a loss of $<$ 0.180 on the Kaggle test set in the final leaderboard ~\cite{challenge}.
The state-of-the-art 3D-CNN, PENet~\cite{huang2020penet}, achieved a mean AUC of around 0.85 for detecting PE. Our study-level PE prediction performance is similar to PENet; however, unlike PENet, our network provides slice-level predictions as well as predictions for associated PE attributes. There are a few limitations to our approach. The CAMs are able to detect blood clots as informative regions with high certainty in unilateral PE. However, in bilateral PE, we can observe that the some discriminative regions are not precisely from the PE regions. This will be investigated in our future work. Additionally
, we will evaluate the efficacy of an end-to-end network and radiomic features~\cite{singh2021radiomics,khorrami2020changes} for the prediction tasks. Refining the radiologic PE diagnostic paradigm through the use of a two-stage network may enable radiologists to reduce inherent human error and also decrease patient morbidity and mortality.
\subsubsection{Acknowledgment:} Dimitris Samaras was partially supported by the Partner University Fund, the SUNY2020 Infrastructure Transportation Security Center, and a gift from Adobe.
\bibliographystyle{splncs04}
\bibliography{mybibliography}

\end{document}